\documentclass[aps,prl,reprint,superscriptaddress]{revtex4-2}

\usepackage{graphicx,epstopdf,siunitx,xspace,amssymb,dsfont,bm,amsmath,float}\usepackage{braket}
\usepackage{chngcntr}

\usepackage{physics}
\usepackage{hyperref}
\usepackage[T1]{fontenc}

\newcommand{\UD}[0]{\ensuremath{\ket{\uparrow \downarrow}} }
\newcommand{\DU}[0]{\ensuremath{\ket{\downarrow \uparrow}} }

\newcommand{\mr}[1]{\ensuremath{\mathrm{#1}}}
\newcommand{\dbz}[0]{\ensuremath{\mr{\Delta}B_z}\xspace}
\newcommand{\eps}[0]{\ensuremath{\epsilon}\xspace}

\DeclareSIUnit \electronvolt {eV}

\begin{document}

\title{Exchange interaction in gate-defined quantum dots beyond the Hubbard model}
\author{Alexander Willmes}
\thanks{These two authors contributed equally}
\affiliation{JARA-Institute for Quantum Information, RWTH Aachen University and Forschungszentrum Jülich GmbH, D-52074 Aachen, Germany}
\author{Patrick Bethke}
\thanks{These two authors contributed equally}
\affiliation{JARA-Institute for Quantum Information, RWTH Aachen University and Forschungszentrum Jülich GmbH, D-52074 Aachen, Germany}
\author{M. Mohamed El Kordy Shehata}
\affiliation{IMEC, Kapeldreef 75, 3001 Leuven, Belgium}
\author{George Simion}
\affiliation{IMEC, Kapeldreef 75, 3001 Leuven, Belgium}
\author{M. A. Wolfe}
\altaffiliation{Also at Department of Physics, University of Wisconsin-Madison, Madison, Wisconsin 53706, USA}
\affiliation{JARA-Institute for Quantum Information, RWTH Aachen University and Forschungszentrum Jülich GmbH, D-52074 Aachen, Germany}
\author{Tim Botzem}
\altaffiliation{Also at School of Electrical Engineering and Telecommunication, UNSW Sydney, Sydney, NSW, Australia}
\affiliation{JARA-Institute for Quantum Information, RWTH Aachen University and Forschungszentrum Jülich GmbH, D-52074 Aachen, Germany}
\author{Robert P. G. McNeil}
\altaffiliation{Also at Center for Quantum Devices, Niels Bohr Institute, University of Copenhagen, Copenhagen 2100, Denmark}
\affiliation{JARA-Institute for Quantum Information, RWTH Aachen University and Forschungszentrum Jülich GmbH, D-52074 Aachen, Germany}
\author{Julian Ritzmann}
\affiliation{Lehrstuhl für Angewandte Festkörperphysik, Ruhr-Universität Bochum, D-44780 Bochum, Germany}
\author{Arne Ludwig}
\affiliation{Lehrstuhl für Angewandte Festkörperphysik, Ruhr-Universität Bochum, D-44780 Bochum, Germany}
\author{Andreas D. Wieck}
\affiliation{Lehrstuhl für Angewandte Festkörperphysik, Ruhr-Universität Bochum, D-44780 Bochum, Germany}
\author{Dieter Schuh}
\affiliation{Institut für Experimentelle und Angewandte Physik, Universität Regensburg, D-93040 Regensburg, Germany}
\author{Dominique Bougeard}
\affiliation{Institut für Experimentelle und Angewandte Physik, Universität Regensburg, D-93040 Regensburg, Germany}
\author{Hendrik Bluhm}
\altaffiliation{bluhm@physik.rwth-aachen.de}
\affiliation{JARA-Institute for Quantum Information, RWTH Aachen University, D-52074 Aachen, Germany}
\affiliation{ARQUE Systems GmbH, D-52074 Aachen, Germany}

\begin{abstract}
A quantitative description of the exchange interaction in quantum dots is relevant for modeling gate operations of spin qubits. By measuring the amplitude and frequency of exchange-driven qubit state oscillations, we measure the detuning dependence of the exchange coupling in a GaAs double quantum dot over three orders of magnitude. Both 1D and 3D full configuration interaction simulations can replicate the observed behavior. Extending a Hubbard model by including excited states increases the range of detuning where it provides a good fit, thus elucidating the underlying physics.
\end{abstract}
	
\maketitle

\paragraph{Introduction.\textemdash}
Many spin qubit implementations rely on the Heisenberg exchange interaction for gate operations. Exchange coupling is a very convenient control resource because it can be electrically controlled by modifying the tunnel coupling or detuning between adjacent quantum dots or donor sites, and is so far the most successful approach for high-fidelity two-qubit gates \cite{xue_highfTQG, Noiri_highfTQG, Mills_highfTQG, Watson_processor, petit_tqg, takeda_tqg, zajac_tqg}. For qubits encoded in several spins, it can also be used for all-electrical single-qubit control \cite{diVincenzo_ExchOnlyTheory, hrl_highfExchOnly, Burkhard-Rev23}.

Experimentally, the use of exchange coupling is well established with recent results showing two-qubit gate fidelities above the error correction threshold \cite{xue_highfTQG,Noiri_highfTQG,Mills_highfTQG}. Understanding the quantitative dependence of the exchange coupling strength $J$ on gate voltages and electric potentials is relevant for aspects like designing optimal control pulses, modeling the sensitivity of qubits to charge noise and achieving a high on-off ratio as required for high-fidelity multi-qubit operations. Yet, such a quantitative understanding is surprisingly incomplete.

Widely used in the literature \cite{Burkhard-Rev23,XuedongHu-Rev00,Reed16,HRLFCI20,Burkhard-FH,Srinivasa-FH} to model the exchange interaction is the Fermi-Hubbard model (FH), due to its simplicity and analytic nature.
For a detuning $\epsilon$ larger than a few tunnel couplings away from the charge transition, the FH model predicts \cite{Reed16, Burkhard-Rev23}
\begin{align}
    J(\epsilon) \propto \frac{1}{\epsilon},
\end{align}
which significantly deviates from experimental observations \cite{Petta_exchExp, Laird_exchExp, Dial_exchExp, Watson_processor, Fogarty18}. In experiments, the exchange has been empirically found to follow
\begin{align}
    J(\epsilon) \sim J_0 \exp \left(\frac{\epsilon}{\epsilon_0} \right)
\end{align}
over narrow ranges of detuning, however a quantitative theory for the parameters $\epsilon_0$ and $J_0$ is still missing.
This exponential dependence can be reproduced by introducing an exponentially detuning-dependent tunnel coupling $t$ \cite{Fogarty18}, but the necessary change in $t$ is larger than 1D simulations support and is experimentally plausible. Additional terms, for example to include effects like triplet hybridization, can be added to the FH model to increase agreement in certain areas \cite{XuedongHu-Rev00}, however the overall discrepancy remains.
The most general and comprehensive approach are 3D full configuration interaction (FCI) models that solve the two-electron Hamiltonian within the effective mass approximation numerically and can include the complete electrostatic potential seen by the electrons \cite{HRLFCI20,Burkhard-Rev23,KordyFCI}. While FCI models can capture effects that the Fermi-Hubbard model might not, FCI models provide little qualitative understanding and require significant computational power. 
Here, we experimentally characterize exchange over a large range of detunings by complementing a standard Ramsey method with the simultaneous extraction of the amplitude and frequency of coherent oscillations of the qubit state while varying the magnetic field gradient. Using this dataset, we compare a simple FH, an extended FH, a 3D FCI as well as a simplified 1D FCI model, which only considers the degree of freedom along the inter-dot axis. We find that extending the simple FH model is useful over a significantly wider regime but still fails to capture exchange over the full detuning range. The 3D FCI model, on the other hand, matches the data well and can be reduced to 1D without losing predictive power if gauged appropriately.

\begin{figure}
    \centering
    \includegraphics[width=\columnwidth]{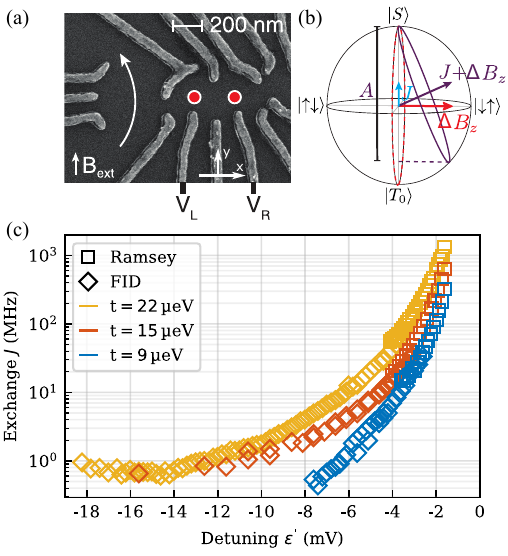}
    \caption{
        \label{fig:data}
        \text{(a)}~Electron micrograph of an identical device. Red circles indicate the dot positions. Gate voltages $V_i$ can be pulsed to change the double dot state on a nanosecond time scale. Reflectance from a nearby sensor dot in an rf tank circuit senses the double dot charge state. We define the x-axis along the inter-dot direction, the z-axis along the out-of-plane direction and y-axis perpendicular to both.
        \text{(b)}~Bloch sphere for the $S$-$T_0$ qubit. Exchange $J$ and nuclear magnetic field gradient \dbz drive rotations around the $z$-/$x$-axis. A finite $J$ tilts the rotation axis, so that the observed frequency is increased and the oscillation amplitude is decreased.
        \text{(c)}~Exchange interaction $J$ measured as a function of detuning $\epsilon^\prime$ using a combination of the Ramsey and combined amplitude-frequency technique. $\epsilon^\prime$ is offset to be zero at the $(2,0)$-$(1,1)$ charge transition. The middle of the $(1,1)$ charge region is situated at approximately $\epsilon^\prime = \SI{-15}{\milli \volt}$. As expected from symmetry, $J(\eps^\prime)$ levels off to a minimum value in the middle of the $(1,1)$ charge region.
    }
\end{figure}

We use a two-electron spin qubit in a double quantum dot formed in a two-dimensional electron gas (2DEG). The 2DEG resides in a GaAs based modulation-doped heterostructure with a $\SI{35}{\nano\meter}$ AlGaAs undoped setback and $\SI{50}{\nano\meter}$ AlGaAs:Si, capped with a $\SI{5}{\nano\meter}$ GaAs:Si protection layer. The electrostatic gating is defined by electron beam lithography (Fig.\,\ref{fig:data}(a)).
The qubit is defined in the $m_z=0$ subspace spanned by $S = (\UD{} - \DU{}) / \sqrt{2}$ and $T_0 = (\UD{} + \DU{}) / \sqrt{2}$, where the first (second) arrow indicates the spin state of the electron in the left (right) quantum dot. The fully polarized triplet states $\ket{T_+} = \ket{\uparrow\uparrow}$ and $\ket{T_-} = \ket{\downarrow\downarrow}$ are split off by the external magnetic field $B_\mr{ext}=\SI{200}{\milli\tesla}$.
In the regime where the qubit states have mostly $(1,1)$-character, with $(N,M)$ indicating the number of electrons in the left ($N$) and right ($M$) quantum dot, the qubit can be described by the Hamiltonian
\begin{align}
    H = \frac{1}{2}\dbz\sigma_x + \frac{1}{2}{J(\eps)}\sigma_z
    \label{eq:hamiltonian}
\end{align}
in the $S$-$T_0$ basis, where $\dbz$ is the difference in magnetic field between left and right dot along the $z$-direction.

Operations on the qubit are performed by rapidly changing the detuning $\eps$, i.e., the energy difference between the left and right quantum dot, by applying a voltage detuning $\eps^\prime=V_L-V_R$, which can be related to $\eps$ using the lever arm $\alpha$ (See Appendix \ref{sec:supl:leverarm}).
This modifies the exchange interaction $J(\eps)$ and induces rotations around the $z$-axis on the Bloch sphere. $\dbz$, which arises mostly from nuclear spins, can be controlled using dynamic nuclear polarization (DNP) \cite{HendrikDNP}.

\paragraph{Exchange interaction measurement.\textemdash}
To characterize the exchange $J$ as a function of detuning $\eps$ all the way from the $(2,0)$-$(1,1)$ charge transition to the middle of the $(1,1)$ charge region, we use a combination of two techniques. Near the charge transition, where $J \gg \dbz$, coherent exchange oscillations around the $z$-axis give access to $J(\eps)$ using a Ramsey-like pulse sequence \cite{Petta_exchExp}. 
The qubit is prepared in the $\UD$ spin state by adiabatically reducing $J(\epsilon)$ to a value where $J \ll \dbz$, pulsed rapidly to the target detuning $\eps_{\text{rot}}$ and evolves freely for some time $\tau$, followed by read out along the $x$-axis.
$J(\eps_{\text{rot}})$ can be extracted from the frequency of the resulting decaying oscillation $A e^{-\tau/T_2^*} \cos(\omega\tau)+c$ (see Fig.\,\ref{fig:ext:ramsey}). However, since 
\begin{align}
\label{eq:omega}
    \omega = \sqrt{J^2+\dbz^2}  \mr{,}
\end{align}
$\dbz$ can make a relevant contribution to $\omega$, especially if $J\lesssim \dbz$. Therefore, we perform an additional free induction decay (FID) experiment, where the qubit is initialized in $\ket{S}$ before pulsing to some detuning $\eps_{\text{ref}}$ where $J(\epsilon_{\text{ref}})\ll \dbz$, giving rise to coherent oscillations around the $x$-axis (dashed circle in Fig.\,\ref{fig:data}(b), see appendix \ref{sec:ext:ramsey} for the corresponding pulse shapes of the Ramsey and FID experiment). 
We repeatedly sweep $\dbz$ by employing DNP and perform multiple repetitions of the FID sequence directly followed by the Ramsey sequence at every $\dbz$ value to extract $J$ from $\omega(\dbz)$.

Due to the increasingly weak $J$-dependence of $\omega$ for $J<\dbz$ and the fact that $\dbz$ cannot be stabilized at arbitrarily small values, this measurement technique has a lower bound in $J$ of $5$ to \SI{10}{\mega \hertz}.
To access smaller values of $J$, we analyze both the amplitude and the frequency of the FID experiment carried out at variable detuning $\eps_{\text{FID}}$.
For $J(\eps_{\text{FID}}) \sim \dbz$, the rotation axis is tilted away from the $x$-axis and the frequency increases according to Equation \ref{eq:omega}, while at the same time the amplitude of the measured oscillations of the qubit state is reduced (continuous, tilted circle in Fig.\,\ref{fig:data}(b)). By sweeping $\dbz$ and extracting the resulting frequency and amplitude variation of the FID, we obtain $J(\eps_{\text{FID}})$ using a simple fit model. 
This approach is particularly useful for systems without single-spin control such as $S$-$T_0$ qubits with a controllable $\dbz$. For details, see Appendix \ref{sec:ext:ampfreq}.

By combining the Ramsey and the amplitude-frequency method, we map out $J(\eps)$ over a wide $\eps$-range for different values of the inter-dot tunnel coupling $t$ (Fig.\,\ref{fig:data}(c)).
As additional input for modeling $J(\eps)$, we measure the charging energy $U$, the ground state (singlet) tunnel coupling $t_S$($=:t$), the exited (triplet) tunnel coupling $t_T$, which is in general different from $t_S$, as well as the singlet-triplet splitting $\Delta_{ST}$ and singlet excited-singlet splitting $\Delta_{SS^*}$, which are given by the distance in detuning between the respective charge transitions. Details are provided in Appendix \ref{sec:supl:tS_tT_meas}.

\paragraph{FCI Models.\textemdash}
Having a comprehensive dataset of the detuning dependence of the exchange interaction at hand, we turn to comparing it to various theoretical models. A model is most useful for predicting the qubit behavior if it reproduces the data, uses parameters that are accessible based on experiments or the potential seen by the electrons, and is computationally inexpensive to evaluate.

The most general approach is a 3D FCI model which solves the two-electron Hamiltonian
\begin{align}
    H = \sum_i^{1,2} \frac{-\hbar^2}{2m^*}\nabla_i^2 + V(\mathbf{r}_i)+\frac{e^2}{4\pi\varepsilon_0\varepsilon_r|\mathbf{r}_1-\mathbf{r}_2|}
\end{align}
where $i$ denotes the two electrons, $\varepsilon_r$ is the relative permittivity and $m^*$ is the effective electron mass.
A simplified 1D version only considers the degree of freedom along the inter-dot axis, which makes it computationally less involved and reduces the number of parameters. To account for the reduced dimensionality and to avoid divergence of the model, we introduce a Coulomb-cutoff length $l_c$ such that the Coulomb interaction for the 1D model is given by
\begin{align}
    E_C = \frac{e^2}{4\pi\varepsilon_0\varepsilon_r} \frac{1}{\sqrt{(x_1-x_2)^2+l_c^2}}\mr{.}
\end{align}
We assume a quartic dependence of the confinement potential in the $x$-direction of the form 
\begin{align}
    V = V_0 \left( (x/x_0)^4 - 2 (x/x_0)^2 \right) + \gamma x
\end{align}
with minima at $\pm x_0$, a barrier height of $V_0$ and a linear term with slope $\gamma$ to detune the potential.
For the 1D model, the energy spectrum as a function of detuning $\gamma$ can then be found with moderate computational effort using exact numerical diagonalization after discretization.\\
For the 3D FCI, model we assume an approximately symmetric shape of the quantum dots in the 2D plane with a quadratic $y$-dependence and self-consistently solve the Poisson-Schrödinger equation to determine the potential of the heterostructure in $z$-direction.
From this potential, the energy spectrum is then obtained using a FCI calculation following Shehata \textit{et al.}\,\cite{KordyFCI}.

In the following, we focus on the measured curve with $t=\SI{22}{\micro\electronvolt}$ as it is the most extensive dataset of the three.
We calibrate each model's detuning to compare it to the experimental data using the slope of the ground-state energy as a function of $\gamma$ far from the avoided crossing and reference all curves to the charge transition.
From the simulated energy diagrams we determine $U$, $\Delta_{ST}$, $\Delta_{SS^*}$, and $t_S$ and $t_T$ as widths and offsets between various avoided crossings to compare them to the experimental values.
As the 1D model is computationally cheaper, we use it to fit $J(\eps)$ to the experiment by varying the inter-dot distance $2 x_0$ and the barrier height $V_0$ such that $t_S$, $t_T$ and $\Delta_{ST}$ match the experimental values.
\begin{figure}
    \centering
    \includegraphics[width=\columnwidth]{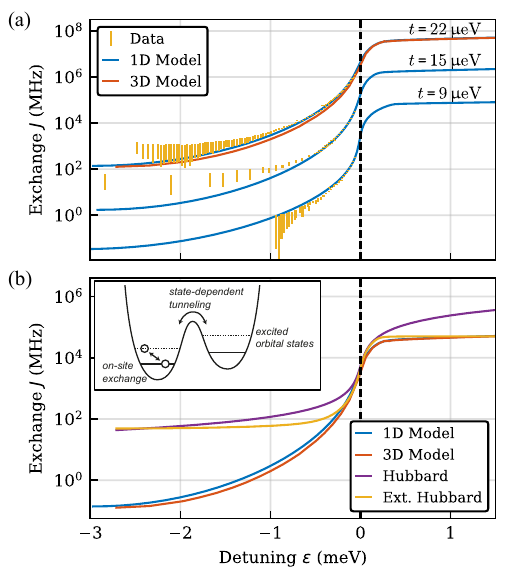}
    \caption{
    \text{(a)} Comparison of the measured $J(\eps)$ with the output of the numerical models. The different curves are labeled with the respective tunnel couplings and offset in $J$ for visual clarity. We use the 3D model to gauge the Coulomb-cutoff length $l_c$ of the 1D model to $l_c = \SI{21.85}{\nano\meter}$ using the curve with $t=\SI{22}{\micro\electronvolt}$ and employ this 1D model to fit the other two measurements.
    Using the experimentally determined lever arm we find good qualitative agreement between model and experiment. For $t=\SI{22}{\micro\electronvolt}$ the potential parameters are $x_0 = \SI{107.5}{\nano\meter}$, $V_0 = \SI{1.575}{\milli \electronvolt}$ and a comparison of extracted parameters with the experiment are shown in Tab.\,\ref{tab:1dfit} (for all parameters of other two measurements see Appendix \ref{sec:supl:other_exchange_fits}).
    The detuning axis is referenced to the $(2,0)$-$(1,1)$ charge transition.
    \text{(b)} Common models with parameters extracted from the exact 1D model. The range over which the extended FH model captures the $J(\eps)$ behavior is much extended, especially in the range of low detuning.
    Inset: Diagram for the additional terms in the extended FH model.
    \label{fig:fits1d}}
\end{figure}

The cutoff length $l_c$ was chosen such that the resulting $J(\epsilon)$ and parameters match those from the 3D FCI calculation in the parameter region of interest. The discrepancies in $J(\epsilon)$ as well as $\Delta_{ST}$ and  $\Delta_{SS^*}$ between the two models primarily arise from a difference in calibration that slightly stretches the detuning axis $\eps$ of the 1D model relative to the 3D model.
For both models we find very good agreement with the experimental parameters (Table \ref{tab:1dfit}) and qualitative agreement on the same order of magnitude as the experimentally measured exchange (Fig.\,\ref{fig:fits1d}(a)). Using the 1D model with the same $l_c$ to fit the other two measured curves yields similarly good agreement, indicating that the gauge of the cutoff length remains sensible for different confinement configurations in $x$-direction (see Appendix \ref{sec:supl:other_exchange_fits}). 
Note that to obtain this match, we had to shift the data by $\SI{0.175}{\milli\electronvolt}$ in $\eps$, relative to the model, for which the $\eps=0$-point was determined from a separate charge stability measurement.
This discrepancy may result from a charge rearrangement in the device between the charge stability measurement and the start of the Ramsey measurement, which seems plausible since a comparable shift of $\sim \SI{0.1}{\milli\electronvolt}$ is visible between the Ramsey and the amplitude-frequency data at around $\eps=\SI{-0.4}{\milli\electronvolt}$ in Fig.\,\ref{fig:fits1d}(a).
The simulated exchange matches well close to the charge transition, but it increasingly diverges from the data points towards the center of $(1,1)$. A possible explanation for this deviation is the oversimplified form of the used potential that cannot fully capture the real quantum dots' shape and specifically neglects any asymmetry that might arise from disorder. 
Both screening by gates and a difference in confinement strength may additionally contribute to the discrepancy in $U$ seen in Tab.\,\ref{tab:1dfit}. Due to the curves being referenced and fitted by parameters close to the charge transition, these effects likely have the largest influence further away in the symmetric configuration ($\eps \sim -U$).

The FCI model and, after gauging $l_c$, also the 1D model have two free parameters, the inter-dot distance $2x_0$ and the barrier height $V_0$. 
The parameters that we extract from the experiment are $t_S$ and $t_T$, which depend on the barrier shape, $\Delta_{ST}$, $U$, which depends on the shape of the potential in the center of the dot. We also measure the energy splitting between the ground and excited state singlet $\Delta_{SS^*}$, which depends on the confinement strength, albeit for a different dot tuning (see Appendix \ref{sec:supl:pulsedgate_spec}). $2x_0$ and $V_0$ are reasonable free parameters to fit the model as the tunnel couplings primarily depend on the shape of the potential in $x$-direction. Note, however, that the excited state energies and $U$ also strongly depend on the lateral confinement, which is not fitted in our procedure for simplicity.
The fitted values for the barrier height and inter-dot distance, $V_0=\SI{1.575}{\milli\electronvolt}$ and $2x_0=\SI{215}{\nano\meter}$, are plausible with respect to a nominal dot distance of $\SI{200}{\nano\meter}$ in the device layout. In combination with the good agreement between the above mentioned parameters as extracted from the experiment and the model (Table \ref{tab:1dfit}) with only two fitting parameters, this shows a good consistency of the model with the experiment.
\begin{table}
    \centering
    \begin{tabular}{lcccc}
        & Experiment & 3D FCI & 1D model & ext. FH \\
        \hline
        $U$ (\si{\milli \electronvolt}) & \num{2.3} & \num{2.7} & \num{3.0} & \num{2.7} \\
        $t_S$ (\si{\micro \electronvolt}) & \num{22.0} & \num{22.0} & \num{21.8} & \num{22.0} \\
        $t_T$ (\si{\micro \electronvolt}) & \num{29.9} & \num{29.2} & \num{29.7} & \num{29.2} \\
        $\Delta_{ST}$ (\si{\milli \electronvolt}) & \num{0.19} & \num{0.21} & \num{0.22} & \num{0.21} \\
        $\Delta_{SS^*}$  (\si{\milli \electronvolt}) & \num{\sim 0.43} & \num{1.11} & \num{1.20} & \num{1.11} \\
        $x$ (\si{\micro \electronvolt}) & & & & 0.01
    \end{tabular}
    \caption{
        Comparison of theory and experiment values for characteristic parameters for Fig.\,\ref{fig:fits1d}
        \label{tab:1dfit}
    }
\end{table}

\paragraph{Hubbard Model.\textemdash}
While the FCI models are suitable for a quantitative description, they provide little insight on the origin of the deviation between the data and the single-orbital Fermi-Hubbard model. To fill this gap, we also consider an extended Fermi-Hubbard model, which increases the detuning range around the charge transition for which a good agreement is obtained.
Starting from a simple three-state Fermi-Hubbard model
\begin{align} 
\label{eq:simpleFH}
H = \begin{pmatrix}
        \frac{\epsilon}{2} & 0 & 0 \\
        0 & \frac{\epsilon}{2} & t \\
        0 & t & U -\frac{\epsilon}{2}
    \end{pmatrix}
\end{align}
in the $T(1,1)$, $S(1,1)$, $S(2,0)$ basis, we extend the orbital basis to include both the $(2,0)$ and $(0,2)$ configuration as well as all singly-excited orbital states.
This inclusion of excited states is motivated as follows. As the standard Fermi-Hubbard model includes no excited orbital states in each dot, there is no charge transition of the triplet states. 
However, experimentally, the triplets also exhibit a charge transition into the (0,2) configuration less than $\SI{1}{\milli\electronvolt}$ from the singlet charge transition, which is relevant within the considered detuning range. Away from the charge transition, the hybridization of the triplet states reduces the exchange splitting compared to the case without triplet-transition. Since excited states see a lower potential barrier, their tunnel coupling is stronger, as seen in Table \ref{tab:1dfit}.
We do not consider doubly excited states, which would lead to a $16 \times 16$ Hamiltonian, because they are physically not needed for the triplet charge transition and their effect is negligible.

Therefore allowing only for singly-excited states and spin-preserving tunneling, we arrive at a $12 \times 12$ Hamiltonian that separates into two orbital sub-Hamiltonians $H_{S}$ and $H_{T}$ (Appendix \ref{sec:ext:extH})  with a symmetrized basis $\{\ket{LL},\ket{LR},\ket{LL^*},\ket{LR^*},\ket{RR},\ket{L^*R},\ket{RR^*}\}$ and an anti-symmetrized basis $\{\ket{LR},\ket{LL^*},\ket{LR^*},\ket{L^*R},\ket{RR^*}\}$ for the singlet and triplet spin configuration respectively, where $L(R)$ denotes the ground orbital state in the left(right) quantum dot and $L^*(R^*)$ denotes excited orbital states.
The extended model contains as parameters the state-dependent tunnel couplings, $t_S=\bra{LL}H_S\ket{LR}$ and $t_T=\bra{LL^*}H_T\ket{LR}$, the ground state on-site interaction $U=\frac{1}{2}(\bra{LL}H_S\ket{LL}+\bra{RR}H_S\ket{RR})$, the direct remote exchange $x=\bra{LR}H_S\ket{LR}$, the direct on-site exchange $\chi_2=\frac{1}{2}\bra{LL^*}(H_S-H_T)\ket{LL^*}$, and the orbital energy $\hbar\omega=\bra{LR^*}H_T\ket{LR^*}$. Note that model parameters can equally be related to other combinations of the matrix elements so that the expressions given reflect a somewhat arbitrary choice.
Values for these terms can be measured independently on the experiment or estimated from numerical and analytical calculations. The direct on-site exchange and orbital energy can be calculated from the experimentally accessible splittings $\Delta_{ST}$ (see Appendix \ref{sec:supl:tS_tT_meas}) and $\Delta_{SS^*}$ (see Appendix \ref{sec:supl:pulsedgate_spec}) as $\chi_2=\frac{1}{2}(\Delta_{SS^*}-\Delta_{ST})$ and $\hbar\omega=\frac{1}{2}(\Delta_{SS^*}+\Delta_{ST})$.

\paragraph{Model comparison.\textemdash}
To compare the models (to each other) we use the parameters obtained from the 3D FCI model, wherefore all models are close to each other near the charge transition by design and no free parameters need to be adjusted. Figure \ref{fig:fits1d}(b) shows the resulting curves. Note that the phenomenological exponential model would appear as a straight line.
We see that the extended Fermi-Hubbard model agrees with the numerical calculations over a much wider range in detuning than the single avoided-crossing model, indicating that the addition of both $(2,0)$ and $(0,2)$ and singly-excited states is a step towards a more accurate analytical description. We attribute the remaining divergence towards the center of $(1,1)$ to the eigenstates and thus tunnel couplings being implicitly assumed as constant in the extended FH model. In reality, a significant change is expected when the detuning is varied on the scale of both the inter-dot barrier height and level splitting, since the energy of the exact single particle double-dot eigenstates will change significantly in this case, which will in turn affect the wave functions. The model also suffers from the direct remote exchange term $x$ being not directly derivable from the energy spectrum, which therefore needs to be estimated and tuned manually.

\paragraph{Conclusion.\textemdash}
In summary, we characterized the exchange interaction as a function of detuning in a GaAs double quantum dot over the full detuning range by employing a dedicated amplitude-frequency method and compare various models to this extensive dataset.
These results lead us to propose the following approach for modeling the exchange interaction between spin qubits in gate-defined quantum dots.
A simple Hubbard model is generally not recommended, as it is only reliable very close to the charge transition. An extended Hubbard model including excited states can be used over a wider range near the charge transition. A 1D configuration interaction model can be used with reasonable accuracy over the full detuning range and still has a moderate computational complexity. It is suitable for estimating the charge noise sensitivity, modeling crosstalk effects and computing transfer functions in the typical operating regime of exchange-based gate operations.
It has the further advantage that the computed wave functions can be used to estimate the influence of different gates, taking the detailed shape of the potential changes they generate into account.
However, its accuracy for quantitative predictions of the residual exchange in the center of the $(1,1)$ region is limited and needs an initial estimation of the Coulomb-cutoff length.
Although the experimental method relies on special properties of GaAs, the results also apply to other host materials for quantum dots, such as SiGe or Si-MOS.
Extensions to this study could address the question how (well) larger exchange couplings in barrier control mode at symmetric detuning could be modeled with the 1D model as well as if and how the low lying valley states typically encountered in Si-based devices have to be included. Furthermore, it would be interesting to what extent a 2D configuration interaction model represents a useful compromise between the used 3D and 1D FCI models.
\\

\paragraph{Acknowledgments.\textemdash}
This project received funding from Deutsche Forschungsgemeinschaft under Grant No. BL 1197/4-1, the Alfried Krupp von Bohlen and Halbach Foundation, and the Excellence Initiative of the German federal and state governments.
J.R., A.L., and A.D.W. acknowledge gratefully support of DFG-TRR160,  BMBF--Q.Link.X  16KIS0867, and the DFH/UFA  CDFA-05-06.
Sample fabrication was carried out at the Helmholtz Nano Facility (HNF) at the Forschungszentrum J\"ulich \cite{albrecht_hnf_2017}.
\\
\paragraph{Data availability.\textemdash}
The data that support the findings of this article are openly available \cite{datalink}.
\\
\paragraph{Author contributions.\textemdash}
Molecular-beam-epitaxy growth of the sample was carried out by J.R., A.L. and A.D.W.. T.B., P.B., and R.P.G.M. set up the experiment. R.P.G.M. fabricated the sample. P.B. conducted the experiment. P.B. and A.W. analyzed the data. A.W. and M.M.E.K.S carried out numerics. A.W. prepared the manuscript. The extended FH model was developed by M.A.W., H.B. and P.B. The 3D FCI model was developed by M.M.E.K.S. and G.S.

\appendix

\setcounter{section}{0}
\renewcommand{\thesection}{\Alph{section}}

\makeatletter
\renewcommand{\section}[1]{
    \refstepcounter{section}
    \par\bigskip
    \begin{center}
        \textbf{Appendix \thesection: #1}
    \end{center}
    \par\bigskip
}
\makeatother

\counterwithin{figure}{section}
\counterwithin{table}{section}
\setcounter{figure}{0}
\setcounter{table}{0}

\section{Amplitude-Frequency Method}
\label{sec:ext:ampfreq}
\begin{figure}
    \centering
    \includegraphics[width=0.99\columnwidth]{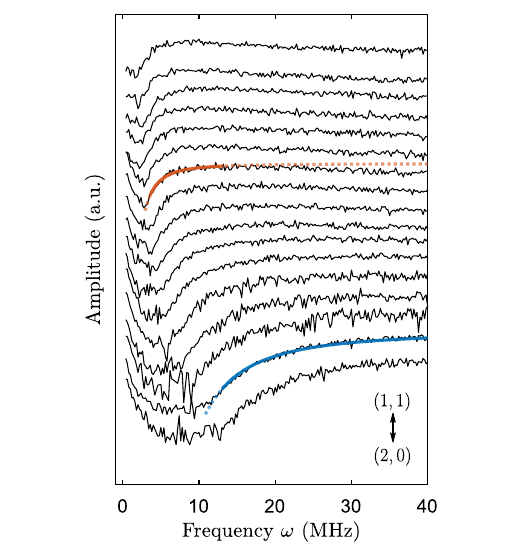}
    \caption{
        \label{fig:ext:ampfreq}
        Measured oscillation amplitude as a function of its frequency using the FID pulse for different pulse depths $\eps$ at $t=\SI{22}{\micro\electronvolt}$. The two colored traces exemplary show fits of Eq.\,\ref{eq:ampfreq}, which are used to extract $J(\eps)$ shown in figure \ref{fig:data}(c) of the main text. The fit range is chosen manually to exclude the noise floor at small $\omega$ and the drop off at large $\omega$, as indicated by the dashed lines
    }
\end{figure}
For small but non-zero values of $J$ the rotation axis of the FID is slightly tilted, as stated in the main text. The frequency of the measured oscillations increases while the amplitude is reduced to
\begin{align}
\label{eq:ampfreq}
    A(\omega) = A_0 \frac{\omega^2 - J^2}{\omega^2}
\end{align}
(Fig.\,\ref{fig:data}(b)), which depends on the observed frequency $\omega$. $A_0$ is the full $S$-$T_0$ contrast.

To measure the frequency and amplitude we Fourier-analyze the time trace of a single FID sequence (Appendix Figure \ref{fig:supl:ampfreq}(a)) and extract the frequency as the bin with maximum power and the amplitude by numerically integrating around this maximum ($\pm \SI{2}{\mega \hertz}$) (Appendix Figure \ref{fig:supl:ampfreq}(c)).
Since the nuclear magnetic field gradient $\dbz$, which arises mostly from nuclear spins, significantly varies even within scanlines, we analyze each repetition individually rather than averaging repetitions of a scanline to reduce the effect of these fluctuations.

By sweeping $\dbz$ back and forth across zero using DNP \cite{HendrikDNP} while measuring frequency and amplitude of the FID, we can extract $A(\omega)$ at a certain detuning $\eps_0$.
To do so, the acquired amplitude-frequency pairs are averaged over $\SI{200}{\kilo\hertz}$-wide frequency bins to give $A(\omega)$ (Appendix Figure \ref{fig:supl:ampfreq}(d)-(e)).
The exchange at detuning $\eps_0$ is then given by the x-intercept, when fitting Equation \ref{eq:ampfreq} to the measured $A(\omega)$ (Fig.\,\ref{fig:ext:ampfreq}).

For large $\dbz$ (large $\omega$ in Fig.\,\ref{fig:ext:ampfreq}) the readout contrast is reduced because of an increased relaxation rate of the metastable triplet state in the readout region \cite{BarthelMetastableT}. While this effect is not included in the fit model, it is negligible in the examined region of small $\dbz$ as the fit is restricted up to the point of maximum amplitude.
The lower bound in $J$ is given by the $\dbz$ fluctuations during a single repetition of the experiment, as the fit model assumes the precession frequency $\omega$ to be constant.
Experimentally, a trade-off for the acquisition time of one repetition has to be made: Shorter individual repetitions limit $\delta\dbz$ and thus increase visibility. However, reducing the number of pulses in the pulse sequence increases noise, and shorter maximum separation times decreases spectral resolution as well as access to low frequencies.

\section{Ramsey and FID Measurement}
\label{sec:ext:ramsey}
\begin{figure}
    \centering
    \includegraphics[width=0.99\columnwidth]{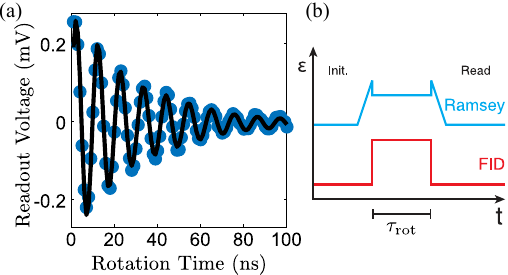}
    \caption{
        \label{fig:ext:ramsey}
        \text{(a)}~Fit $A\text{cos}(\omega)e^{t/T_2^*}$ to the Ramsey measurement results ($J=\SI{91}{\mega\hertz}$). Using an exponential decay for the oscillations provides a better fit to the data a Gaussian decay, which may be related to non-ergodicity effects of the system. \text{(b)}~Pulse sequences used in this work (offset in $\eps$). For the Ramsey sequence $\UD$ is adiabatically prepared and read out to observe $z$-rotations. For the FID sequence a singlet is prepared and rapidly separated to observe $S-T_0$ oscillations.
    }
\end{figure}

For the Ramsey measurement the qubit is prepared in the $\UD$ spin state by adiabatically reducing $J(\epsilon)$ to a value where $J \ll \dbz$, pulsed rapidly to the target detuning $\eps_{rot}$ and evolves freely around the $z$-axis for some time $\tau_{rot}$, followed by read out on the $\UD$-$\DU$ axis (Fig.\,\ref{fig:ext:ramsey}(b)).
Figure \ref{fig:ext:ramsey}(a) shows a representative Ramsey measurement of the exchange oscillations.

\section{Extended Hubbard Hamiltonian}
\label{sec:ext:extH}
The complete extended Hubbard Hamiltonian of the symmetric singlet subspace is given by
\begin{widetext}
\begin{align} 
\label{eq:ext:SingletHam}
H_S = \begin{pmatrix}
        U-\epsilon & -t_S & 0 & -t_T & 0 & 0 & 0 \\
        -t_S & x & -t_T & 0 & -t_S & 0 & -t_T \\
        0 & -t_T & U+\chi_2+\hbar\omega-\epsilon & -t_T & 0 & -t_S & 0 \\
        -t_T & 0 & -t_T & x+\hbar\omega & 0 & 0 & -t_S \\
        0 & -t_S & 0 & 0 & U+\epsilon & -t_T & 0 \\
        0 & 0 & -t_S & 0 & -t_T & x+\hbar \omega & -t_T \\
        0 & -t_T & 0 & -t_S & 0 & -t_T & U+\chi_2+\hbar\omega+\epsilon
    \end{pmatrix}
\end{align}
with the basis states $\{\ket{LL},\ket{LR},\ket{LL^*},\ket{LR^*},\ket{RR},\ket{L^*R},\ket{RR^*}\}$ and for the anti-symmetric triplet subspace by
\begin{align} 
\label{eq:ext:TripletHam}
H_T = \begin{pmatrix}
        -x & -t_T & 0 & 0 & t_T \\
        -t_T & U-\chi_2+\hbar\omega-\epsilon & -t_T & t_S & 0 \\
        0 & -t_T & \hbar\omega & 0 & -t_S \\
        0 & t_S & 0 & \hbar\omega & t_T \\
        t_T & 0 & -t_S & t_T & U-\chi_2+\hbar\omega+\epsilon 
    \end{pmatrix}
\end{align}
\end{widetext}
with basis states $\{\ket{LR},\ket{LL^*},\ket{LR^*},\ket{L^*R},\ket{RR^*}\}$.
Note that we assume the same tunnel coupling $t_T$ for all singly excited states as well as the same on-site energy $U$ and direct remote exchange term $x$ irrespective of whether the states are singly excited or not.

\section{Independent tunnel coupling measurement}
\label{sec:supl:tS_tT_meas}
\begin{figure}[h!]
    \centering
    \includegraphics[width=.99\columnwidth]{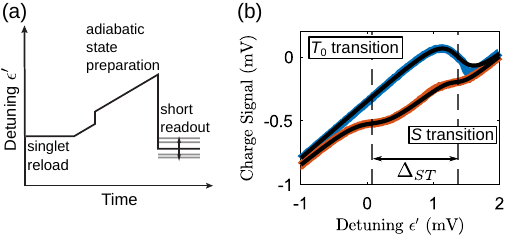}
    \caption{
        \label{fig:ext:tS_tT_meas}
        \text{(a)} Pulse sequence for the excited-state tunneling measurement $t_T$. The $\ket{\uparrow\downarrow}$ state is prepared adiabatically and pulsed back rapidly to some detuning where the charge state is read out for $\SI{100}{\nano\second}$ to $\SI{200}{\nano\second}$, which is short compared to the lifetime of the prepared state. \text{(b)} Charge signal of the singlet ground state (blue) and the $\ket{\uparrow\downarrow}$ state (orange) measurement. The width of the transition labeled ${S}$ (${T_0}$) transition gives $t_S$ ($t_T$) while the distance between both transitions corresponds to the singlet-triplet splitting $\Delta_{ST}$. The linear background reflects the direct effect of the gates on the charge sensor.
    }
\end{figure}

To measure the ground state (singlet) tunnel coupling $t_S$($=:t$) we slowly sweep over the $(2,0)$-$(1,1)$ transition while recording the charge sensor response. As the tunnel coupling of the exited (triplet) state $t_T$ is generally different and not accessible with this method, we employ Pauli-Spin blockade by adiabatically initializing in $\UD$ and rapidly pulsing to variable detunings near the charge transition (Fig. \ref{fig:ext:tS_tT_meas}(a)). For each pulse, the charge sensor response of the initialized metastable state is recorded at the target detuning on a timescale that is short compared to the relaxation time.  Averaging over many repetition leads to a data set as shown in Figure \ref{fig:ext:tS_tT_meas}(b), from which $t_T$ (as well as $t_S$) can be extracted. 
It also yields the singlet-triplet splitting $\Delta_{ST}$ via the distance in detuning between both, singlet and metastable triplet, charge transitions. We determine the values of $t_S$ and $t_T$ by fitting an avoided crossing model of the form
\begin{align}
    A + B\cdot(\eps^\prime-\eps_S^\prime)+ \sum_{i=S,T} C_i\frac{\eps^\prime-\eps_i^\prime}{\sqrt{(\eps^\prime-\eps_i^\prime)^2+4w_i^2}},
\end{align}
to the data (Figure \ref{fig:ext:tS_tT_meas}(b)). Here, $\eps_{S,T}^\prime$ represent the positions of the two transitions, $w_{S,T}$ the tunnel couplings in voltage, and $A$ and $B$ correspond to the SET background signal and the direct coupling of the gates to the SET, respectively. We use $A$, $B$, $C_i$, $\eps_i^\prime$, and $w_i$ as fit parameters.
The width and position of the ground state charge transition agree well between both measurements.
The transitions are in principle broadened by thermal fluctuations, however in our typical operating regime, $t_{S,T}$ is well above $k_\mr{B} T$ and we therefore neglect the thermal broadening.

\section{Lever Arm Measurement}
\label{sec:supl:leverarm}
\begin{figure}[h]
    \centering
    \includegraphics[width=.99\columnwidth]{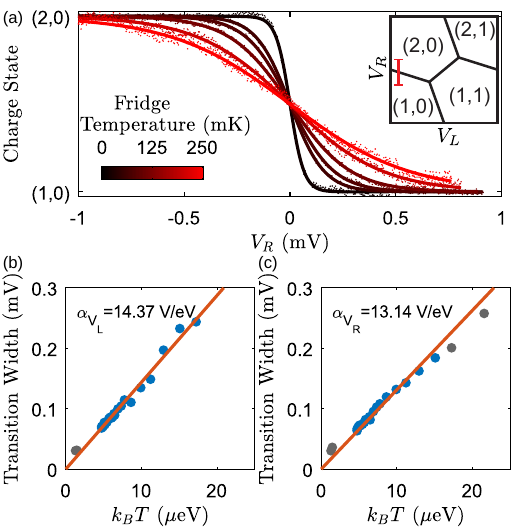}
    \caption{
        \label{fig:supl:leverarm}
        \text{(a)} Charge state occupation as a function of $V_R$ and mixing chamber temperature. The electrons thermalize with the substrate and the reservoir charge transition is thermally broadened. (Inset) Sketch of the charge stability diagram. Sweep range indicated in red. \text{(b-c)} Width of the $(1,0) \leftrightarrow (2,0)/(1,0) \leftrightarrow (1,1)$ charge transition as a function of thermal energy. The linear behavior indicates an electron temperature below $\SI{50}{\milli\kelvin}$, the slope gives the lever arm $\alpha$ for the $V_L$ and $V_R$ gates.
        }
\end{figure}
For comparison between the theoretical models and the measured exchange curves, we transform the experimental gate voltages along the detuning axis $\eps^\prime$ into energy units, i.e.\, $\eps = \frac{\eps^\prime}{\alpha}$, where $\alpha$ is the lever arm, whose measurement is explained in the following.

The thermal broadening of charge transitions between a dot and its lead can be used to characterize the electron temperature $T_e$ of the sample \cite{DiCarloMethod} using
\begin{align}
    A+B V_i&+C \tanh{\left(\frac{V_i-V_{i,0}}{2k_BT_e}\right)}\mr{.}
\end{align}
With typical tunneling times of order 10 ns, the leads from the dots to the reservoir are opaque compared to the electron temperature so that the width of the charge transition is dominated by thermal broadening.
By controlling the temperature of the sample with external heaters in the dilution refrigerator, we can measure the transition width $w$ as a function of temperature and determine the lever arm $\alpha$, as demonstrated in Fig.\,\ref{fig:supl:leverarm}.
The temperature then acts as an absolute energy scale used to gauge the gate voltages via the lever arm. The values for the lever arms of the RF gates are in the expected range set by similar devices. The lever arm along the detuning axis can be calculated from the single-gate lever arms as
\begin{equation}
    \alpha_\eps^{-1}=\alpha_{V_L}^{-1}+\alpha_{V_R}^{-1}
\end{equation}
where $V_L$ and $V_R$ correspond to the plunger gates labeled in Figure \ref{fig:data}(a) of the main text.
Because of the limited thermometer calibration range we can only give an upper limit of roughly $\SI{50}{\milli\kelvin}$ to the electron temperature.

\section{Amplitude-frequency data}
\label{sec:supl:ampfreq}
\begin{figure}[h]
    \centering
    \includegraphics[width=0.99\columnwidth]{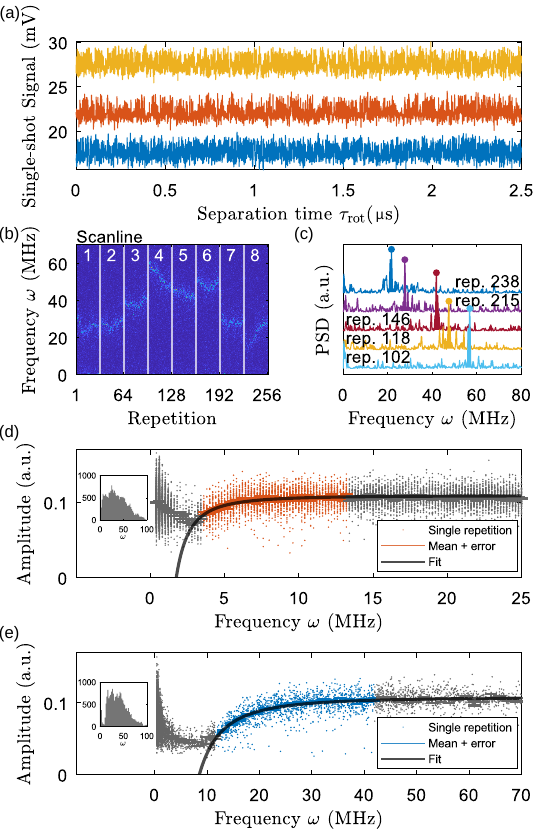}
    \caption{
        \label{fig:supl:ampfreq}
        \text{(a)}~Time domain signals of the $\Delta B_z$ FID sequence. Each data point shows the average of the readout signal over the $\SI{7}{\micro\second}$ readout window of one pulse.
        \text{(b)}~Example PSD for subsequent repetitions of the FID pulse measurement (panel (a)). Each scanline (white lines) consists of 32 repetitions taking $\SI{560}{\milli\second}$ in total after which DNP pulses are applied for $\SI{1}{\second}$. Even within scanlines $\dbz$ fluctuates significantly, indicating that single-shot treatment is necessary.
        \text{(c)}~Single traces from (b). The frequency is extracted as the maximum power of the PSD and the amplitude by numerically integrating over $\SI{\pm 2}{\mega\hertz}$ bins around the peak power.
        \text{(d-e)}~Example fits of the oscillation amplitude vs. frequency $\omega$ for $\eps^\prime=\SI{-6.41}{\milli\volt}$ ($\eps^\prime=\SI{-10.01}{\milli\volt}$) colored in blue (orange) with same color code as in Fig. \ref{fig:data}(c) of the main text. Fitting the model in Eq.\,\ref{eq:ampfreq} to $A(\omega)$ gives $J(\eps)$ as the $x$-axis intercept. The fit range is chosen manually to not include the noise floor at small $\omega$. The inset histogram shows how often each frequency was sampled.
    }
\end{figure}

Figure \ref{fig:supl:ampfreq} exemplary shows the workflow from raw data to exchange fit for the amplitude frequency method.

\section{Additional exchange fits}
\label{sec:supl:other_exchange_fits}

To fit the two curves with tunnel coupling $t=\SI{9}{\micro\electronvolt}$ and $t=\SI{15}{\micro\electronvolt}$ we use the 1D model with the Coulomb-cutoff length $l_c=\SI{21.85}{\nano\meter}$ gauged from $t=\SI{22}{\micro\electronvolt}$ as described in the main text. This seems justified since in order to vary the tunnel coupling of the double quantum dot predominantly the inter-dot barrier is modified. Therefore, the most significant change in potential takes place in the $x$-direction, which does not influence $l_c$. This rationale is supported by the good agreement of the 1D model to both measurements. The extracted parameters for these two curves is given in Table \ref{tab:supl:1dfitOther}. Again, a detuning-offset of $\SI{84}{\micro\electronvolt}$ and $\SI{81}{\micro\electronvolt}$ had to be applied to the curves with $t=\SI{9}{\micro\electronvolt}$ and $t=\SI{15}{\micro\electronvolt}$, respectively, to align the data with the simulation. Due to a technical issue, the measurement of the triplet tunnel coupling for $t=\SI{9}{\micro\electronvolt}$ could not be completed and only the singlet tunnel coupling was used for fitting. $U$ and $\Delta_{SS^*}$ were only measured once (see Appendix \ref{sec:supl:pulsedgate_spec}) and are therefore the same as in the main text.
\\
\\
\\
\\
\begin{table*}[h]
    \centering
    \begin{tabular}{lccccc}
        & $U$ (\si{\milli \electronvolt}) & $t_S$ (\si{\micro \electronvolt}) & $t_T$ (\si{\micro \electronvolt}) & $\Delta_{ST}$ (\si{\milli \electronvolt}) & $\Delta_{SS^*}$  (\si{\milli \electronvolt}) \\
        \hline
         $t=\SI{9}{\micro \electronvolt}$\\
        \hline
        Experiment & $\sim 2.3$ & \num{9.1} & - & - & $\sim0.43$ \\
        1D model & \num{3.2} & \num{9.1} & \num{14.7} & \num{0.36} & \num{1.36} \\
        \hline
        $t=\SI{15}{\micro \electronvolt}$\\
        \hline
        Experiment & $\sim 2.3$ & \num{14.6} & \num{17.5} & \num{0.19} & $\sim 0.43$ \\
        1D model & \num{3.0} & \num{13.9} & \num{19.1} & \num{0.19} & \num{1.15}
    \end{tabular}
    \caption{
        Comparison of theory and experiment values for characteristic parameters of the other two measurements of Fig.\,\ref{fig:fits1d}(a). $U$ and $\Delta_{SS^*}$ were only measured once and are therefore the same as in the main text. The Coulomb-cutoff length was chosen the same for all three fits, i.e. $l_c=\SI{21.85}{\nano\meter}$. Other parameters for the 1D model are $x_0=\SI{106.3}{\nano\meter}$ and $V_0=\SI{2.14}{\milli\electronvolt}$ and $x_0=\SI{115.5}{\nano\meter}$ and $V_0=\SI{1.62}{\milli\electronvolt}$ for $t=\SI{9}{\micro\electronvolt}$ and $t=\SI{15}{\micro\electronvolt}$ respectively.
        \label{tab:supl:1dfitOther}
    }
\end{table*}

\section{Time Domain Pulsed-Gate Spectroscopy}
\label{sec:supl:pulsedgate_spec}
\begin{figure*}[h]
    \centering
    \includegraphics[width=0.99\textwidth]{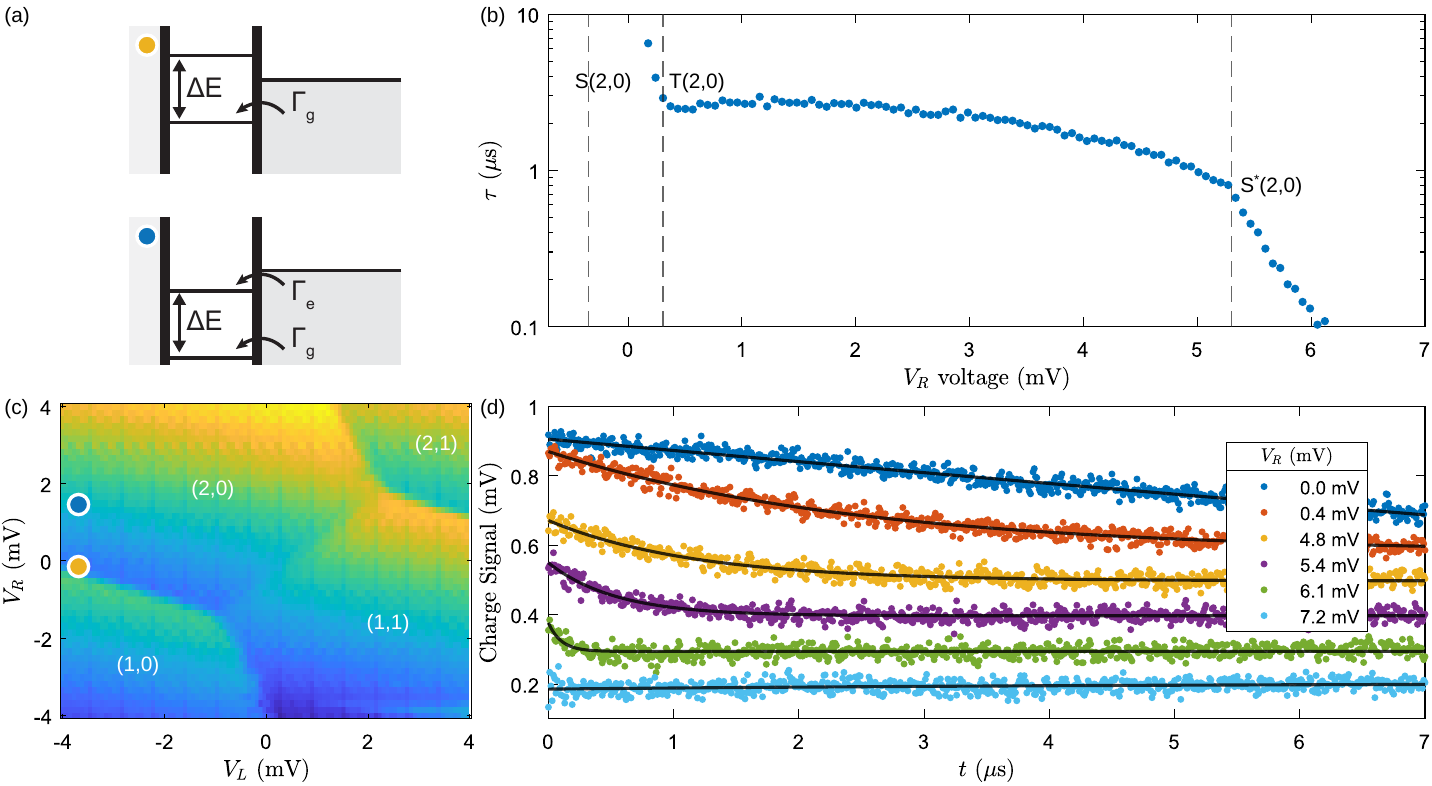}
    \caption{
        \label{fig:supl:pulsedgate_spec}
        \text{(a)} Simplified energy scheme. The total tunneling rate $\Gamma$ depends on the number of states available below the Fermi energy. Once the plunge depth reaches the orbital energy spacing, an additional state becomes available with an additional tunneling channel $\Gamma_e$, reducing the charge relaxation time.
        \text{(b)} Charge relaxation time $\tau$ as a function of plunge depth. The position of ground state charge transition is known form the charge stability diagram. The triplet charge transition and the excited orbital charge transition are visible by abrupt changes in the tunneling times.
        Before the T(2,0) state becomes available, $\tau$ is much longer than the pulse length, so that fits are unreliable.
        \text{(c)} Charge stability measurement. The pulsed experiment was carried out at an plunger voltage of $V_L=\SI{-4}{\milli\volt}$. The blue and yellow circles indicate the corresponding situation in (a)
        \text{(d)} Relaxation time traces at different plunge depths. The curves are offset for clarity.
        }
\end{figure*}

To investigate the excited energy levels of the double dot system, we perform time domain pulsed-gate spectroscopy. For this the double dot is initialized in the $(1,0)$ charge configuration and a square voltage pulse with varying plunge depth $V_R$ pulses the double dot into the $(2,0)$ regime. Due to our high readout bandwidth, we are able to look at the tunneling times directly by recording the charge sensor signal in a time-resolved manner. The time it takes to draw in an additional electron depends on the number of states that are available and the tunneling time to the leads. When a new state becomes energetically available, the tunneling time changes abruptly as visible in Figure \ref{fig:supl:pulsedgate_spec}(b). In the measurement we can see the $T(2,0)$ and the $S^*(2,0)$ states becoming available. From Figure \ref{fig:supl:pulsedgate_spec}(b) we can extract the singlet-triplet splitting as $\Delta_{ST} \approx \SI{0.05}{\milli\electronvolt}$ and the singlet-excited singlet splitting as $\Delta_{SS^*} \approx \SI{0.43}{\milli\electronvolt}$. The singlet-triplet splitting deviates from the value in the main text, because the dot-reservoir coupling has to be tuned into a range that enables this type of measurement and hence differs significantly from the tuning in the main text. Therefore the values measured here can only be taken as an order of magnitude estimation.

\bibliography{arxiv_refs.bib}

@PREAMBLE{
 "\providecommand{\noopsort}[1]{}" 
 # "\providecommand{\singleletter}[1]{#1}%" 
}

@article{Fogarty18,
  title = {Integrated silicon qubit platform with single-spin addressability, exchange control and single-shot singlet-triplet readout},
  author = {M. A. Fogarty and K. W. Chan and B. Hensen and W. Huang and T. Tanttu and C. H. Yang and A. Laucht and M. Veldhorst and F. E. Hudson and K. M. Itoh and D. Culcer and T. D. Ladd and A. Morello and A. S. Dzurak},
  journal = {Nat Commun},
  volume = {9},
  pages = {4370},
  year = {2018},
  month = {Oct},
  doi = {10.1038/s41467-018-06039-x}
}

@article{HRLFCI20,
  title = {Resonant exchange operation in triple-quantum-dot qubits for spin–photon transduction},
  author = {Andrew Pan and Tyler E Keating and Mark F Gyure and Emily J Pritchett and Samuel Quinn and Richard S Ross and Thaddeus D Ladd and Joseph Kerckhoff},
  journal = {Quantum Science and Technology},
  volume = {5},
  pages = {034005},
  year = {2020},
  month = {May},
  doi = {10.1088/2058-9565/ab86c9}
}

@article{XuedongHu-Rev00,
  title = {Hilbert-space structure of a solid-state quantum computer: Two-electron states of a double-quantum-dot artificial molecule},
  author = {Xuedong Hu and S. Das Sarma},
  journal = {Phys. Rev. A},
  volume = {61},
  pages = {062301},
  year = {2000},
  month = {May},
  doi = {10.1103/PhysRevA.61.062301}
}

@article{Reed16,
  title = {Reduced Sensitivity to Charge Noise in Semiconductor Spin Qubits via Symmetric Operation},
  author = {M.D. Reed and B.M. Maune and R.W. Andrews and M.G. Borselli and K. Eng and M.P. Jura and A.A. Kiselev and T.D. Ladd and S.T. Merkel and I. Milosavljevic and E.J. Pritchett and M.T. Rakher and R.S. Ross and A.E. Schmitz and A. Smith and J.A. Wright and M.F. Gyure and and A.T. Hunter},
  journal = {Phys. Rev. Lett.},
  volume = {116},
  pages = {110402},
  year = {2016},
  month = {March},
  doi = {10.1103/PhysRevLett.116.110402}
}

@article{Burkhard-Rev23,
  title = {Semiconductor spin qubits},
  author = {Guido Burkard and Thaddeus D. Ladd and Andrew Pan and John M. Nichol and and Jason R. Petta},
  journal = {Rev. Mod. Phys.},
  volume = {95},
  pages = {025003},
  year = {2023},
  month = {June},
  doi = {10.1103/RevModPhys.95.025003}
}

@article{Burkhard-FH,
  title = {Three-electron spin qubits},
  author = {Maximilian Russ1 and Guido Burkard},
  journal = {J. Phys.: Condens. Matter},
  volume = {29},
  pages = {393001},
  year = {2017},
  month = {August},
  doi = {10.1088/1361-648X/aa761f}
}

@article{Srinivasa-FH,
  title = {Entangling distant resonant exchange qubits via circuit quantum electrodynamics},
  author = {V. Srinivasa and J. M. Taylor and Charles Tahan},
  journal = {Phys. Rev. B},
  volume = {94},
  pages = {205421},
  year = {2016},
  month = {Nov},
  doi = {10.1103/PhysRevB.94.205421}
}

@article{KordyFCI,
  title = {Modeling semiconductor spin qubits and their charge noise environment for quantum gate fidelity estimation},
  author = {M. Mohamed El Kordy Shehata and George Simion and Ruoyu Li and Fahd A. Mohiyaddin and Danny Wan and Massimo Mongillo and Bogdan Govoreanu and Iuliana Radu and Kristiaan De Greve and Pol Van Dorpe},
  journal = {Phys. Rev. B},
  volume = {108},
  pages = {045305},
  year = {2023},
  month = {July},
  doi = {10.1103/PhysRevB.108.045305}
}

@article{HendrikDNP,
  title = {Enhancing the Coherence of a Spin Qubit by Operating it as a Feedback Loop That Controls its Nuclear Spin Bath},
  author = {Hendrik Bluhm and Sandra Foletti and Diana Mahalu and Vladimir Umansky and Amir Yacoby},
  journal = {Phys. Rev. Lett.},
  volume = {105},
  pages = {216803},
  year = {2010},
  month = {November},
  doi = {10.1103/PhysRevLett.105.216803}
}

@article{BarthelMetastableT,
  title = {Relaxation and readout visibility of a singlet-triplet qubit in an Overhauser field gradient},
  author = {C. Barthel and J. Medford and H. Bluhm and A. Yacoby and C. M. Marcus and M. P. Hanson and A. C. Gossard},
  journal = {Phys. Rev. B},
  volume = {85},
  pages = {035306},
  year = {2012},
  month = {Janurary},
  doi = {10.1103/PhysRevB.85.035306}
}

@article{DiCarloMethod,
  title = {Differential Charge Sensing and Charge Delocalization in a Tunable Double Quantum Dot},
  author = {L. DiCarlo and H. J. Lynch and A. C. Johnson and L. I. Childress and K. Crockett and C. M. Marcus and M. P. Hanson and A. C. Gossard},
  journal = {Phys. Rev. Lett},
  volume = {92},
  pages = {226801},
  year = {2004},
  month = {June},
  doi = {10.1103/PhysRevLett.92.226801}
}

@article{Petta_exchExp,
  title = {Coherent Manipulation of Coupled Electron Spins in Semiconductor Quantum Dots},
  author = {J. R. Petta and A. C. Johnson and J. M. Taylor and E. A. Laird and A. Yacoby and M. D. Lukin and C. M. Marcus and M. P. Hanson and A. C. Gossard},
  journal = {Science},
  volume = {309},
  pages = {2180-2184},
  year = {2005},
  month = {September},
  doi = {10.1126/science.1116955}
}

@article{Dial_exchExp,
  title = {Charge Noise Spectroscopy Using Coherent Exchange Oscillations in a Singlet-Triplet Qubit},
  author = {Dial, O. E. and Shulman, M. D. and Harvey, S. P. and Bluhm, H. and Umansky, V. and Yacoby, A.},
  journal = {Phys. Rev. Lett.},
  volume = {110},
  issue = {14},
  pages = {146804},
  numpages = {5},
  year = {2013},
  month = {Apr},
  publisher = {American Physical Society},
  doi = {10.1103/PhysRevLett.110.146804},
  url = {https://link.aps.org/doi/10.1103/PhysRevLett.110.146804}
}

@article{Watson_processor,
  title = {A programmable two-qubit quantum processor in silicon},
  author = {T. F. Watson and S. G. J. Philips and E. Kawakami and D. R. Ward and P. Scarlino and M. Veldhorst and D. E. Savage and M. G. Lagally and Mark Friesen and S. N. Coppersmith and M. A. Eriksson and L. M. K. Vandersypen},
  journal = {Nature},
  volume = {555},
  pages = {633-637},
  year = {2018},
  month = {February1},
  doi = {10.1038/nature25766}
}

@article{Laird_exchExp,
  title = {Effect of Exchange Interaction on Spin Dephasing in a Double Quantum Dot},
  author = {Laird, E. A. and Petta, J. R. and Johnson, A. C. and Marcus, C. M. and Yacoby, A. and Hanson, M. P. and Gossard, A. C.},
  journal = {Phys. Rev. Lett.},
  volume = {97},
  issue = {5},
  pages = {056801},
  numpages = {4},
  year = {2006},
  month = {Jul},
  publisher = {American Physical Society},
  doi = {10.1103/PhysRevLett.97.056801},
  url = {https://link.aps.org/doi/10.1103/PhysRevLett.97.056801}
}

@article{xue_highfTQG,
    author = {Xiao Xue and Maximilian Russ and Nodar Samkharadze and Brennan Undseth and Amir Sammak and Giordano Scappucci and Lieven M. K. Vandersypen},
    title = {Quantum logic with spin qubits crossing the surface code threshold},
    journal = {Nature},
    volume = {601},
    pages = {343-347},
    year = {2022},
    month = {January},
    doi = {10.1038/s41586-021-04273-w}
}

@article{Noiri_highfTQG,
    author = {Akito Noiri and Kenta Takeda and Takashi Nakajima and Takashi Kobayashi and Amir Sammak and Giordano Scappucci and Seigo Tarucha},
    title = {Fast universal quantum gate above the fault-tolerance threshold in silicon},
    journal = {Nature},
    volume = {601},
    pages = {338-342},
    year = {2022},
    month = {January},
    doi = {10.1038/s41586-021-04182-y}
}

@article{Mills_highfTQG,
    author = {ADAM R. Mills and CHARLES R. Guinn and MICHAEL J. Gullans and ANTHONY J. Sigillito and MAYER M. Feldman and ERIK Nielsen and JASON R. Petta},
    title = {Two-qubit silicon quantum processor with operation fidelity exceeding 99\%},
    journal = {Sci. Adv.},
    volume = {8},
    issue = {14},
    pages = {eabn5130},
    year = {2022},
    month = {April},
    doi = {10.1126/sciadv.abn5130}
}

@article{hrl_highfExchOnly,
    author = {Aaron J. Weinstein and Matthew D. Reed and Aaron M. Jones and Reed W. Andrews and David Barnes and Jacob Z. Blumoff and Larken E. Euliss and Kevin Eng and Bryan H. Fong and Sieu D. Ha and Daniel R. Hulbert and Clayton A. C. Jackson and Michael Jura and Tyler E. Keating and Joseph Kerckhoff and Andrey A. Kiselev and Justine Matten and Golam Sabbir and Aaron Smith and Jeffrey Wright and Matthew T. Rakher and Thaddeus D. Ladd and Matthew G. Borselli},
    title = {Universal logic with encoded spin qubits in silicon},
    journal = {Nature},
    volume = {615},
    pages = {817-822},
    year = {2023},
    month = {February},
    doi = {10.1038/s41586-023-05777-3}
}

@article{diVincenzo_ExchOnlyTheory,
    author = {D. P. DiVincenzo and D. Bacon and J. Kempe and G. Burkard and K. B. Whaley},
    title = {Universal quantum computation with the exchange interaction},
    journal = {Nature},
    volume = {408},
    pages = {339-342},
    year = {2000},
    month = {November},
    doi = {10.1038/35042541}
}

@article{petit_tqg,
    author = {L. Petit and H. G. J. Eenink and M. Russ and W. I. L. Lawrie and N. W. Hendrickx and S. G. J. Philips and J. S. Clarke and L. M. K. Vandersypen and M. Veldhorst},
    title = {Universal quantum logic in hot silicon qubits},
    journal = {Nature},
    volume = {580},
    pages = {355-359},
    year = {2020},
    month = {April},
    doi = {10.1038/s41586-020-2170-7}
}

@article{takeda_tqg,
    author = {Kenta Takeda and Akito Noiri and Takashi Nakajima and Takashi Kobayashi and Seigo Tarucha},
    title = {Quantum error correction with silicon spin qubits},
    journal = {Nature},
    volume = {608},
    pages = {682-686},
    year = {2022},
    month = {August},
    doi = {10.1038/s41586-022-04986-6}
}

@article{zajac_tqg,
    author = {D. M. Zajac and A. J. Sigillito and M. Russ and F. Borjans and J. M. Taylor and G. Burkard and J. R. Petta},
    title = {Resonantly driven CNOT gate for electron spins},
    journal = {Science},
    volume = {359},
    issue = {6374},
    pages = {439-442},
    year = {2017},
    month = {December},
    doi = {10.1126/science.aao596}
}

@article{albrecht_hnf_2017,
    author = {W.Albrecht and J. Moers and B. Hermanns},
    title = {HNF - Helmholtz Nano Facility},
    journal = {JLSRF},
    volume = {3},
    pages = {A112},
    year = {2017},
    month = {September},
    doi = {10.17815/jlsrf-3-158}
}

@article{datalink,
  author       = {A. Willmes and P. Bethke and M.M.E.K. Shehata and G. Simion and M.A. Wolfe and T. Botzem and R.P.G. McNeil and J. Ritzmann and A. Ludwig and A.D. Wieck and D. Schuh and D. Bougeard and H. Bluhm},
  title        = {Data for: Exchange interaction in gate-defined quantum dots beyond the Hubbard model},
  year         = {2025},
  journal      = {Zenodo},
  doi          = {10.5281/zenodo.17722982},
  url          = {https://doi.org/10.5281/zenodo.17722982},
}
\clearpage

\end{document}